\documentstyle[epsfig,mathptm]{mn} 
\newif\ifAMStwofonts
\AMStwofontstrue                % <- comment this out if AMS fonts not available
\DeclareMathVersion{bold}                    % <- seems to be needed with mathptm
\DeclareMathAlphabet{\mathbfit}{OT1}{cmr}{bx}{it}
\SetMathAlphabet\mathbfit{bold}{OT1}{cmr}{bx}{it}
\DeclareMathAlphabet{\mathbfss}{OT1}{cmss}{bx}{n}
\SetMathAlphabet\mathbfss{bold}{OT1}{cmss}{bx}{n}
\ifAMStwofonts
  \ifCUPmtlplainloaded \else
    \DeclareSymbolFont{UPM}{U}{eur}{m}{n}
    \SetSymbolFont{UPM}{bold}{U}{eur}{b}{n}
    \DeclareSymbolFont{AMSa}{U}{msa}{m}{n}
    \DeclareMathSymbol{\upi}{0}{UPM}{"19}
    \DeclareMathSymbol{\umu}{0}{UPM}{"16}
    \DeclareMathSymbol{\upartial}{0}{UPM}{"40}
    \DeclareMathSymbol{\leqslant}{3}{AMSa}{"36}
    \DeclareMathSymbol{\geqslant}{3}{AMSa}{"3E}

    \let\geq=\geqslant 
  \fi
\fi

\newcommand{\apj}{{ ApJ}}
\newcommand{\apjs}{{ApJS}}
\newcommand{\aj}{{ AJ}}

\newcommand{\mnras}{{MNRAS}}

\newcommand{\aap}{{ A\&A}}

\newcommand{\nat}{{ Nature}}
\title
[Cosmological parameters from cluster abundances, CMB and IRAS]
{Cosmological parameters from cluster abundances, CMB and IRAS}
\author[S.L.~Bridle et al.]
{S.L.~Bridle$^1$, V.R.~Eke$^2$, O.~Lahav$^{2,3}$, A.N.~Lasenby$^1$, 
M.P.~Hobson$^1$, S.~Cole$^4$,
\newauthor C.S.~Frenk$^4$ and J.P.~Henry$^5$\\ 
$^1$Astrophysics Group, Cavendish Laboratory,  Madingley Road, 
Cambridge CB3 0HE, UK\\
$^2$Institute of Astronomy, Madingley Road, Cambridge CB3 0HA, UK\\
$^3$Racah Institute of Physics, The Hebrew University, Jerusalem 
91904, Israel\\
$^4$Department of Physics, University of Durham, South Road, Durham
DH13LE\\
$^5$Institute for Astronomy, 2680 Woodlawn Drive, Honolulu, H1 96822, USA\\
}

\date{Accepted ???. Received ???; in original form \today}
\pagerange{\pageref{firstpage}--\pageref{lastpage}}
\pubyear{1999}
\begin{document}
\maketitle
\label{firstpage}
\begin{abstract} 
We combine information on cosmological parameters from cluster
abundances, CMB primordial anisotropies and the IRAS 1.2 Jy galaxy
redshift survey. We take as free parameters the present values of the
total matter density of the universe, $\Omega_{\rm{m}}$, the Hubble
parameter, $h$, the linear theory rms fluctuations in the matter
density within $8 h^{-1}$ Mpc spheres, $\sigma_8$, and the IRAS
biasing factor, $b_{\rm IRAS}$. We assume that the universe is
spatially flat, with a cosmological constant, and that structure
formed from adiabatic initial fluctuations with a Harrison-Zel'dovich
power spectrum (i.e. the primordial spectral index $n=1$). The
nucleosynthesis value for the baryonic matter density $\Omega_{\rm{b}}
= 0.019/h^2$ is adopted. We use the full three- and four- dimensional
likelihood functions for each data set and marginalise these to two-
and one- dimensional distributions in a Bayesian way, integrating over
the other parameters. It is shown that the three data sets are in
excellent agreement, with a best fit point of $\Omega_{\rm{m}} =
1-\Omega_{\Lambda} = 0.36$, $h = 0.54$, $\sigma_8 = 0.74$, and $b_{\rm
IRAS} = 1.08$. This point is within one sigma of the minimum for each
data set alone. Pairs of these data sets have their degeneracies in
sufficiently different directions that using only two data sets at a
time is sufficient to place good constraints on the cosmological
parameters. We show that the results from each of the three possible
pairings of the data are also in good agreement. Finally, we combine
all three data sets 
to obtain marginalised 68 per cent confidence
intervals of $0.30<\Omega_{\rm{m}}<0.43$, $0.48<h<0.59$,
$0.69<\sigma_8<0.79$, and $1.01<b_{\rm IRAS}<1.16$. For the best fit
parameters the CMB quadrupole is $Q_{\rm{rms-ps}}=18.0 \mu$K, the
shape parameter of the mass power-spectrum is $\Gamma=0.15$, the
baryon density is $\Omega_{\rm{b}}=0.066$ and the age of the universe
is $16.7$~Gyr. 
\end{abstract}
%\clearpage

\section{Introduction}
\label{intro}

Combining and comparing information from different types of
observations is a powerful and important tool in cosmology. In this
letter we investigate the agreement between the conclusions drawn from
studies of CMB data, a galaxy redshift survey and the variation of the
cluster abundance with redshift.

The number density of rich galaxy clusters at low redshift depends
strongly on both $\Omega_{\rm{m}}$ and $\sigma_8$, with a weak
dependence on the shape of the power spectrum of fluctuations (Lilje
1992; Bahcall \& Cen 1993; Hanami 1993; White, Efstathiou \& Frenk
1993). Measuring the evolution of the number density with redshift
breaks the degeneracy between these two main parameters (eg. Oukbir \&
Blanchard 1992; Viana \& Liddle 1996; Eke, Cole \& Frenk 1996). The
comparison of the X-ray cluster temperature functions (the number of
clusters per unit volume as a function of the temperature of their
X-ray emitting gas) determined at low (Edge et al. 1990; Henry \&
Arnaud 1991) and high (Henry 1997) redshift is a good way to implement
this test. Henry (1997) first performed this operation, and Eke et
al. (1998) and Viana \& Liddle (1999) have considered additional
systematic uncertainties.  

The power spectrum of the primary anisotropies in the Cosmic Microwave
Background (CMB) depends on many cosmological parameters. The
constraints that can be inferred from current CMB data arise mainly
from the amplitude measured for the fluctuations on large angular
scales and also the position and height of a peak in the power
spectrum at smaller scales. Discussions of the implied cosmological
parameter information, given a range of assumptions, have been most
recently presented by Hancock et al. (1998), Bond \& Jaffe (1998),
Lineweaver (1998), Tegmark (1998), Bartlett et al. (1998) and
Efstathiou et al. (1998). 

While the CMB and cluster abundances depend on the fluctuations 
in mass, galaxy redshift surveys tell us about the distribution 
of luminous matter. It is commonly (and probably naively) assumed that
the distribution of luminous galaxies of a certain type is related to
the underlying matter distribution by the `biasing factor', $b$. We
define this as the ratio of the rms fluctuations in the galaxy number
density to those in the underlying mass. Thus, the amplitude of the
galaxy power spectrum is $b^2$ larger than that of the mass. Under the
assumption that $b$ is scale-independent, the shape of the galaxy
power spectrum is the same as that of the matter power spectrum.
Thus the galaxy power spectrum can be used to constrain the matter
power spectrum shape parameter, $\Gamma$. As the cluster abundances
only weakly constrain $\Gamma$, the addition of galaxy redshift
surveys to the joint analysis helps to fix the shape of the power
spectrum. In addition, the redshift space distortion of the galaxy
distribution due to peculiar velocities yields a measure of the
combination, $\beta=\Omega_{\rm{m}}^{0.6}/b$. The joint analysis with
CMB and cluster abundances allows us to remove the degeneracy in this
combination. Constraints on cosmological parameters from redshift
surveys have been obtained by Fisher, Scharf \& Lahav (1994), Tadros
et al. (1999), Sutherland et al. (1999) and Branchini et al. (1999). 

Many authors have now compared various combinations of data sets at a
range of levels of detail (Bond \& Jaffe 1998; Gawiser \& Silk 1998;
Lineweaver 1998; White 1998; Garnavich et al. 1998; Webster et
al. 1998; Eisenstein, Hu \& Tegmark 1998; Efstathiou et al. 1998 and
Roos \& Harun-or-Rashid 1999). We choose to perform a joint analysis
of CMB, galaxy redshift survey and cluster abundance data sets, since 
they provide complementary constraints on cosmological
parameters. As we demonstrate below, the parameter degeneracies 
arising from these three data sets are relatively
complicated and merit the full likelihood approach adopted in this
paper.  

In Section 2 we introduce the data and outline the methods used to
produce likelihoods for each separate set of observations. We describe
our method of combination and some statistical issues in Section 3 and
present our results in Section 4, discussing them in more detail in
Section 5.

\begin{figure*}
\centerline{\vbox{\epsfig{file=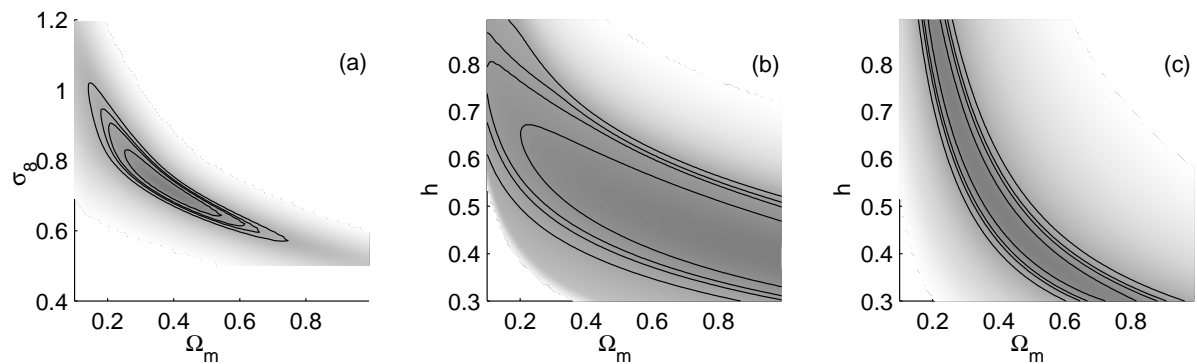,height=16cm, angle=90}}}
\caption
{(a) The cluster abundance likelihood function in the
\{$\Omega_{\rm{m}}$,$\sigma_8$\}-plane, after marginalisation over
$h$. (b) The CMB likelihood function in the \{$\Omega_{\rm{m}}$,
$h$\}-plane, after marginalisation over $\sigma_8$. (c) The IRAS
likelihood function in the \{$\Omega_{\rm{m}}$, $h$\}-plane, after
marginalisation over $\sigma_8$ and $b_{\rm IRAS}$. The contours
denote the 68, 90, 95 and 99 per cent confidence regions. 
\label{eachdata}}
\end{figure*}

\section{Individual data sets}
\label{data}

\subsection{Cluster Abundances}
\label{clab}

A range of different values for cosmological parameters have been inferred
from the evolution of the number density of clusters with redshift
(e.g. Frenk et al. 1990; Carlberg et al. 1997; Bahcall, Fan \& Cen
1997; Blanchard \& Bartlett 1998; Reichart et al. 1999), and in some
cases the quoted uncertainties are sufficiently small to make the
results incompatible. This implies that systematic uncertainties are
important (see Eke et al. 1998 for a discussion). For a joint
likelihood analysis, the resulting confidence limits on the inferred
cosmological parameters are only reliable if the systematic errors
present in the individual analyses are less important than the
statistical ones. For this reason, the cluster abundance is addressed
using the X-ray temperature function, where this is least unlikely to
be the case.  

We use two X-ray flux-limited samples. The first is comprised of the
25 clusters with average redshift 0.05 compiled by Henry \& Arnaud
(1991). We have incorporated the latest temperature measurements from
Ginga and ASCA for these clusters (Henry 1999). The second sample is
comprised of 14 Einstein Extended Medium Sensitivity Survey (EMSS)
clusters with average redshift 0.38 compiled by Henry (1999). The
temperatures of all 14 clusters have been determined with ASCA data
using the most recent calibrations available (Henry 1999). In order to
calculate likelihoods the Press-Schechter (1974) expression for the
abundance of rich clusters as a function of mass and redshift, coupled
with a mass-temperature conversion (White et al. 1993; Eke et
al. 1998) is employed to predict the distribution of cluster redshifts
and temperatures for each model. These are compared with the observed
distributions as discussed by Henry (1997) and Eke et al. (1998),
using the `default' assumptions detailed by Eke et al. (1998).

As discussed in Section 1, the main constraint from cluster abundances
is in the $\sigma_8$, $\Omega_{\rm{m}}$ plane, and this is plotted in
Fig. \ref{eachdata} (a).

\subsection{CMB}
\label{cmb}

Our approach follows that of Hancock et al. (1998) and Webster et
al. (1998). We use the same data as Webster et al., with the addition
of the recent QMAP points (de Oliveira-Costa et al. 1998). This same
data set is used and plotted in Efstathiou et al. (1998). 

The flat band-power method is employed (as described in e.g. Hancock
et al. 1998), although Bartlett et al. (1999) and Bond, Jaffe \& Knox
(1998) find that this approximation widens the error ranges on the
cosmological parameters. The conclusions of this letter would not be
significantly affected by such a change. 

We assume that there are negligible tensor contributions to the CMB
power spectrum, as predicted by most inflation models. We also assume
neglible re-ionisation and a Harrison-Zel'dovich primordial power
spectrum (ie. the primordial spectral index $n=1$) as favoured by
inflation and the CMB data. We note that changing these assumptions
would substantially alter our results. The remaining family of Cold
Dark Matter models are analysed using the Boltzmann code of Seljak \&
Zaldarriaga (1996).  

The CMB COBE data point constrains the large scale temperature
fluctuations well. This converts to a strong constraint on $\sigma_8$ for
each $h$ and $\Omega_{\rm{m}}$. Estimates of the location and height of the
first acoustic peak are also provided by the CMB data, and these
constrain a degenerate combination of $h$ and $\Omega_{\rm{m}}$
(Fig. \ref{eachdata} (b)). Putting these together also places a weak
constraint on $\sigma_8$. 

\subsection{IRAS 1.2 Jy survey}
\label{IRAS}

The data and method used here are exactly as in Fisher et al. (1994)
and Webster et al. (1998). We use a sample of 5313 IRAS galaxies
flux-limited to 1.2 Jy at 60 $\mu$m, selected from the IRAS data base
(Strauss et al. 1990, Fisher 1992), and calculate redshift weighted
spherical harmonics which may be compared with those predicted given a
matter power spectrum. We use Gaussian windows for the redshift
weighting, and take into account the redshift distortion using an
approximation that is correct only in linear theory. Therefore in this
analysis we use only spherical harmonics up to $l=10$ so that the
scales probed are large, and the fluctuations can be assumed to be in
the linear regime. 

It is assumed that the relationship between the galaxy
distribution and the matter distribution can be described by the single
parameter, the biasing factor for the IRAS galaxies, $b_{\rm
IRAS}$. In practice the relationship between the two power spectra may
have many dependencies, including scale dependence. Note that if
$b_{\rm IRAS}$ is actually scale dependent then the information on the
shape of the power spectrum given by the IRAS data would be wrong. However,
if the assumption of scale independent biasing is correct then galaxy
survey data such as the IRAS 1.2 Jy survey place powerful constraints
on the shape of the power spectrum in the scale range examined. The
redshift distortion also provides limits on $\beta_{\rm IRAS} =
\Omega_{\rm{m}}^{0.6}/b_{\rm{IRAS}}$ and hence a constraint on
$\Omega_{\rm{m}}$ for each value of $b_{\rm{IRAS}}$. However, this turns out to
be a relatively weak constraint. Nevertheless, the range of values of
$\Omega_{\rm{m}}$ allowed by IRAS together with CMB or cluster abundances
lies well within the range allowed by IRAS alone. 
Thus, for this paper, the most important constraint from the IRAS data
is that on the shape of the matter power spectrum which, as will be
seen in the next section, manifests itself in the $\Omega_{\rm{m}}$,
$h$ plane, plotted in Fig. \ref{eachdata} (c).

\section{Combining the data sets}
\label{method}

\subsection{Parameters}
\label{pars}

In this letter the three cosmological parameters we vary are: the
present day value of the reduced Hubble constant, 
$h=H_0/(100$~\mbox{$\rm{km} \rm{s}^{-1} \rm{Mpc}^{-1}$}$)$, 
the matter density of the universe in units of the
critical density, $\Omega_{\rm{m}}$, and the matter power spectrum
normalisation parameter, $\sigma_8$, the rms linear fluctuation of matter
density in $8 h^{-1}$ Mpc spheres. The fourth parameter is $b_{\rm IRAS}$,
which should be regarded as a `fudge factor' reflecting our ignorance
on how to relate fluctuations in galaxy counts to fluctuations in
mass. We assume that the universe is flat so that
$\Omega_{\rm{m}}+\Omega_{\Lambda}=1$ as suggested by recent analyses of CMB
data which allow for $\Omega_{\rm{m}}$ and $\Omega_{\Lambda}$ to vary
independently (eg. Lineweaver 1998; Tegmark 1998; Efstathiou et
al. 1998).

To transform the information on the shape of the matter power spectrum
provided by the IRAS and cluster abundance data into constraints in
the parameter space we consider, we use the Bardeen et al. (1986)
approximation formula for the matter power spectrum, parameterised by
the shape parameter, $\Gamma$. We use the approximate expression
provided by Sugiyama (1995), 
\begin{equation}
\Gamma = \Omega_{\rm{m}}\,h \ {\rm exp}\left({-\Omega_{\rm{b}}\left[1 +
{{\sqrt{h/0.5}}\over{\Omega_{\rm{m}}}}\right]}\right) \ 
\label{eqngamma}.
\end{equation}
This, and the CMB power spectrum calculation, require a value for the
baryon density which we take from the nucleosynthesis constraint,
$\Omega_{\rm{b}} h^2=0.019$ (Burles \& Tytler 1998). Our results are
relatively insensitive to this value. 

Using this conservative number of parameters allows a more thorough
examination of the parameter space, and despite our large
number of assumptions, we find remarkable agreement between the data
sets.

\subsection{Statistical issues}
\label{stats}

Comparing results from different data sets enables us to 
shed some light on how well our current picture of cosmology stands
up. This kind of comparison could be performed by simply
comparing the parameter values predicted by different data . 
While this is a start, it does not
take into account the fact that most measurements are sensitive to more
than one parameter. A full examination of the level of agreement between
data sets should include a comparison of likelihood values in
multi-dimensional space. In this letter we have considered the four
free parameters $\Omega_{\rm{m}}$, $\sigma_8$, $h$ and $b_{\rm IRAS}$ and found that there
\emph{is} a region in the four-dimensional space in which the likelihood
functions of each data set are high. This is important to check
before simply multiplying likelihoods, because the final best fit point
could be a bad fit to all the data sets used, even if the one-dimensional marginalised joint likelihood functions for each parameter
look promising.  
The goodness of fit of each data set can be evaluated e.g. 
by $\chi^2$.

\begin{figure}
\centerline{\vbox{\epsfig{file=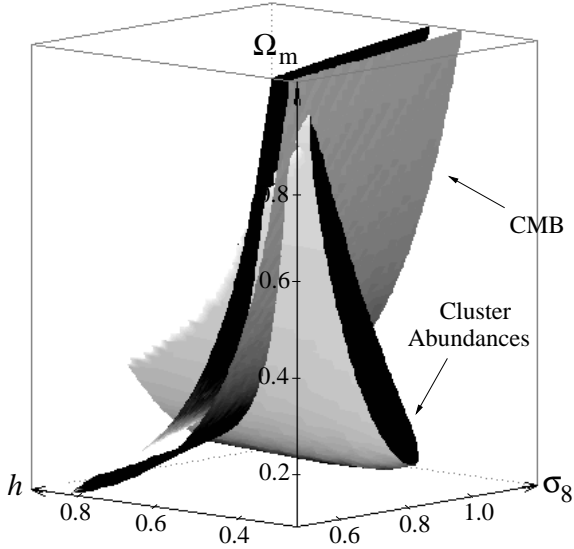,height=8cm}}}
\caption{
CMB and cluster abundance iso-probability surfaces plotted in
three-dimensions, for $\Delta$log(Likelihood)=7.1 (3$\sigma$ for a 3d
Gaussian probability distribution).}
\label{cmbclab3d}
\end{figure}

Combining results is also powerful because the parameter degeneracies from 
each data set may be complementary.
Fig. \ref{cmbclab3d} shows a three-dimensional view of the
orthogonality of the CMB and cluster abundance confidence surfaces. 
The favoured region of parameter space is very
much smaller than the regions of parameter 
space allowed by the single data sets alone, thus yielding much tighter
constraints.
This joint
likelihood function is computed by
calculating likelihoods over a grid in parameter space,
then multiplying them together for the different data sets.

Note that multiplication of the likelihoods for each data set is strictly
correct only in the case where the data sets are \emph{independent}. While
the CMB clearly probes a different part of the universe to either the
IRAS galaxies or the cluster abundance clusters, it is not so clear
whether there is some overlap in the IRAS and low redshift cluster
data. In this letter we assume that these two data sets are
independent, and justify it on the basis that (a) the redshift ranges
do not overlap significantly, 
the low redshift clusters having a median
redshift of 0.05, 
while the IRAS galaxies have a median redshift of 0.02
and (b) because IRAS galaxies are selected in the infra-red,
they tend to be field galaxies and so are not likely to be in the
clusters sampled in the cluster survey. This issue also affects the
CMB analysis, since the flat band power $\chi^2$ analysis is not valid
if the observations used overlap on the sky and in angular scale range
(spherical harmonic $\ell$ range). We have avoided this by using a
set of CMB data points that do not overlap in this way, but the set of
data points chosen could be argued to be somewhat subjective.

We obtain confidence limits on each cosmological parameter by
marginalising over the other parameters. If the parameters are well
enough constrained then the integration limits used for the
marginalisation will not affect the results. The Bayesian
interpretation of the integration limits is that we have `top hat'
priors on each parameter. 
For the integration in the marginalisation we use $0.2 <\sigma_8<1.4$,
$0.3<h<0.9$, $0.1<\Omega_{\rm{m}}<1.0$ and $0.7<b_{\rm IRAS}<2.0$ (in 100, 100, 100, 50
steps respectively).
The parameters are so poorly constrained by any single
data set that these integration limits do affect some
marginalised results. For example, our range of integration in $\sigma_8$
affects the two-dimensional plot of $h$ and $\Omega_{\rm{m}}$
for the CMB data (Fig. \ref{eachdata} (b)), making the contours
foreshorten at low $\Omega_{\rm{m}}$. 
However, we choose our limits to be wide enough that they do not
affect the results when using more than one data set at a time. Thus
the one-dimensional marginalised distributions for pairs of data sets
plotted in Fig. \ref{1ddists} (a) are independent of the priors.
This integration process is computationally costly, and some authors
utilise approximations at this point. For example Lineweaver (1998)
and Tegmark (1998) `project' the likelihood function to one dimension,
taking the maximum value of the likelihood found on varying the other
parameters, rather than integrating over all the values of the
likelihood for all the possible values of the other parameters. This
approximation is correct when the likelihood function is a Gaussian,
but not otherwise.
Note that when we talk about the 68 per cent confidence surface or
contour we mean the (iso-probability) boundary to the region
containing 68 per cent of the probability (given our top hat priors). 

\section{Results}
\label{results}

\begin{figure*}
%\centerline{
\vbox{\epsfig{file=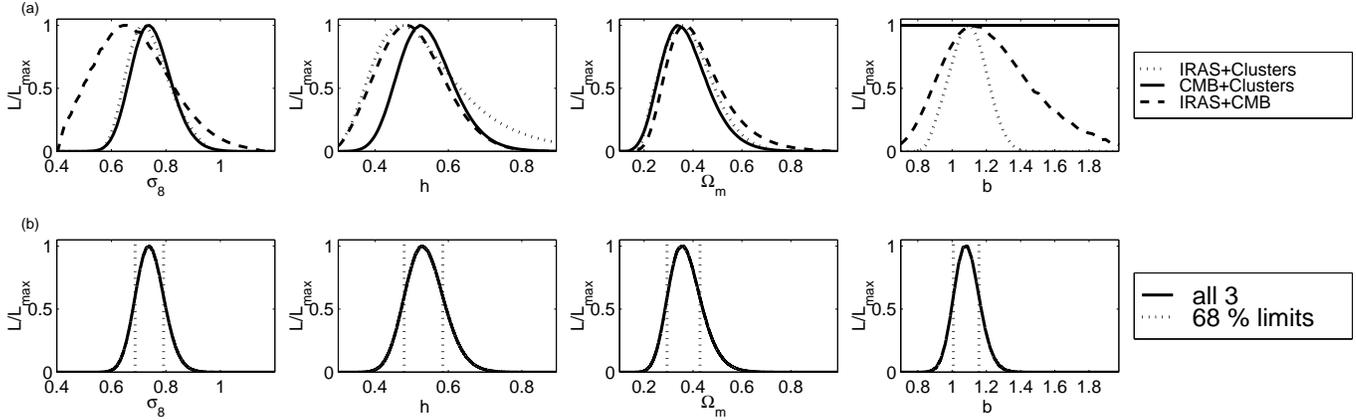,height=18cm, angle=90}}
%}
\caption
{
The 1d marginalised distributions using (a) pairs of data sets and (b)
using all three data sets. The dotted
lines denote 68 per cent confidence limits.
\label{1ddists}}
\end{figure*}

\begin{figure*}
\centerline{\vbox{\epsfig{file=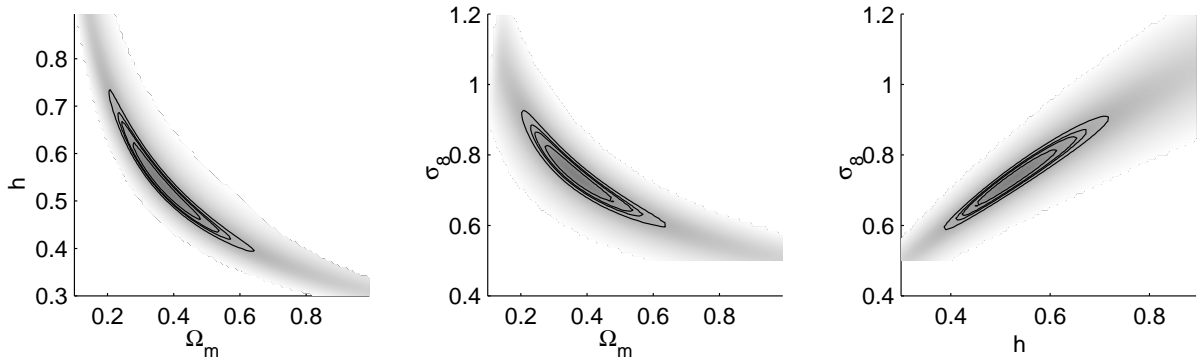,height=16cm, angle=90}}}
\caption{
The likelihood function from combining all three data sets as a function
of pairs of parameters, in each case marginalised over the remaining two
parameters. The contours denote the 68, 90, 95 and 99 per cent
confidence regions. 
\label{all3_2dmarg}}
\end{figure*}

Since it is possible to choose reasonable integration ranges such that
the results using any pair of data sets are independent of the
integration ranges, we may say that just a pair of data sets is
sufficient to constrain properly these four parameters (or three
parameters for CMB and cluster abundances).   

The results for pairs of data sets are shown in Fig. \ref{1ddists}. It
can be seen that these three data sets are in extremely good
agreement, since the parameter values preferred by each pair are
consistent. 
This is particularly apparent for $\Omega_{\rm{m}}$. It is a
result of the fact that all three 68 per cent confidence surfaces intersect in the same
place in three and four dimensions. It was therefore not surprising to find that the
best fit point given in Table 1 is within the 68 per cent confidence surface for each
single data set. 
\begin{table}
\centerline{\vbox{
\begin{tabular}{@{}lcc}
\hline
Parameter & Best fit point & 68 per cent confidence limits\\
\hline
$\Omega_{\rm{m}}$    &$0.36$  &  $0.30 \,<\,\Omega_{\rm{m}}\,<\,0.43$\\
$h$                  &$0.54$  &  $0.48 \,<\,h              \,<\,0.59$ \\
$\sigma_8$           &$0.74$  &  $0.69 \,<\,\sigma_8       \,<\,0.79$\\
$b_{\rm IRAS}$       &$1.08$  &  $1.01 \,<\,b_{\rm IRAS}   \,<\,1.16$\\
\hline
\end{tabular}
}}
\label{bestfit}
\caption{
Parameter values at the joint optimum.
The 68\% confidence limits are shown,
calculated for each parameter by marginalising the likelihood over the other 
variables.}
\end{table}

For the parameter values at the best fit point, the shape parameter
for the mass power spectrum $\Gamma=0.15 \pm 0.02$, the baryonic
density $\Omega_{\rm{b}}=0.066 \pm 0.013$, the CMB quadrupole
$Q_{\rm{rms-ps}}=18.0 \pm 1.0\mu$K, the IRAS distortion parameter
$\beta_{\rm{IRAS}} \equiv \Omega_{\rm{m}}^{0.6}/ b_{\rm IRAS} = 0.50
\pm 0.05$ and the age of the universe is $16.7 \pm 1.0$ Gyr. In each
case, the errors correspond to the estimated 68 per cent marginalised
confidence limits. 

As an indicator of how well the joint optimum fits each data set alone
we calculate the $\chi^2$ value at the joint optimum for each of the
CMB and IRAS data sets, using the method described in Webster et
al. (1998).  
The CMB $\chi^2$ is 22.8 (we use 23 CMB data points) and the IRAS
$\chi^2$ is $142$ (for spherical harmonics up to $l=10$ there are
$120$ degrees of freedom), suggesting that the predicted observations
agree well with the actual observations for the parameters at the
joint optimum.  
Because of the unbinned nature of the cluster abundance analysis, it
is more complicated to find a similar statistic. However, as discussed
by Eke et al. (1998), all of the models considered predict a
relatively even distribution of clusters over the range $0.3<z<0.4$,
but in actual fact there are many more clusters at the low end of this
redshift range than predicted by any of the models considered. This
suggests that the selection function for the clusters was poorly
modelled, inevitably leading to 
a poor goodness of fit to the cluster data, independent of the
parameter values used in the model (given the assumptions of this
paper). 
The number of clusters predicted from the best fitting model is $19.5$
at low redshift and $26$ at high redshift. This may be compared to the
actual numbers observed, $25$ and $14$ respectively, and can be seen
to differ by more than the 1$\sigma$ of $\sqrt{N}$ expected from
Poisson statistics.

The 1d and 2d marginalised distributions given all three data sets are 
also shown in Figs. \ref{1ddists} and \ref{all3_2dmarg}. It can be
seen that adding in a third data set 
tightens the constraints further and the 68 per cent confidence limits
obtained from these 1d distributions are presented in Table 1.

\section{Discussion}
\label{discussion}

The non-trivial and reassuring result is that the marginalised distributions
of the parameters are consistent when single data sets, or any combination
of two data sets, are included. This means that no single experiment is
dominating or pushing the favoured parameter values away from the otherwise 
preferred ones.
It is thus tempting to suggest that systematic uncertainties in the
individual data sets may not be greatly affecting the parameter values that
have been inferred by this analysis.

The strong agreement between pairs of data sets can be understood
qualitatively as follows. The cluster abundance data provide a strong
constraint on $\sigma_8$; the CMB provides a much weaker limit on $\sigma_8$
but one which agrees with the cluster abundance limit. The IRAS data
provides limits on the combination $\sigma_8 b_{IRAS}$ and so
putting this together with the above constraint on $\sigma_8$ produces a
small allowed range of $b_{\rm IRAS}$. The cluster abundance data also favour
just a small range in $\Omega_{\rm{m}}$. The different degeneracy
directions in the $\Omega_{\rm{m}}$,$h$ plane from each of the CMB and IRAS
data produce a slightly weaker limit on $\Omega_{\rm{m}}$, but one which agrees with
the cluster abundance result. Putting together the strong cluster
abundance constraint on $\Omega_{\rm{m}}$ with either the CMB or IRAS limits
in the $h$, $\Omega_{\rm{m}}$ plane produces a constraint on $h$;
alternatively, just the CMB and IRAS data can be used to determine an
allowed range in $h$; and each of these routes to $h$ produces the
same answer.  

Our results agree well with other recent work which
combines data sets to set limits on cosmological parameters (Bond \&
Jaffe 1998, Gawiser \& Silk 1998, Lineweaver 1998, White 1998,
Garnavich et al. 1998, Eisenstein et al. 1998 and Roos \&
Harun-or-Rashid 1999). 

The recent Type 1a supernova results of Perlmutter et al. (1998) imply
that for a flat universe $\Omega_{\rm{m}}=0.28^{+0.09}_{-0.08}$ (1$\sigma$
statistical errors), in good agreement with the values
derived in this letter. Similar results are also found by Riess et al. (1998). 
From gravitational lensing measurements, Falco, Kochanek \& Munoz
(1998) find that $\Omega_{\rm{m}}>0.38$ at 2 sigma, although Chiba and Yoshi
(1999) find that a flat universe with
$\Omega_{\rm{m}}=0.3^{+0.2}_{-0.1}$ is preferred. 
Our value for the Hubble constant, $h = 0.54$, falls at the low end of
the latest distance ladder measurements (Freedman et al. 1998).
The optimal baryon density (given $\Omega_{\rm{b}} h^2 = 0.019$) is
$\Omega_{\rm{b}}=0.066$, and dividing by our best fit value of
$\Omega_{\rm{m}}=0.36$ our optimal baryon fraction is $0.18$
This may be compared to cluster baryon fraction
estimates, for example, White et al. (1993) find $\Omega_{\rm{b}} / \Omega_{\rm{m}}
\geq 0.015 + 0.08 h^{-3/2}$. Inserting our best fit value for
$h$ yields $\Omega_{\rm{b}} /\Omega_{\rm{m}} \geq 0.22$.
Our value for the combination $\sigma_8 \Omega_{\rm{m}}^{0.6} = 0.40$  is lower 
than the one derived from measurements from the peculiar velocity
field $\sigma_8 \Omega_{\rm{m}}^{0.6} \approx 0.8$ (Freudling et al. 1998). It is
closer to the value recently derived from cluster peculiar velocities
by Watkins (1997) of $\sigma_8 \Omega_{\rm{m}}^{0.6} = 0.44^{+0.19}_{-0.13}$.
Reassuringly, the age of the universe we find ($16.7$ Gyr) is greater
than the age of the oldest stars (Chaboyer et al. 1998).

\subsection*{ACKNOWLEDGMENTS}

We would like to thank Gra\c ca Rocha for her work on compilation of the
CMB data set.
SLB, VRE, SC and CSF acknowledge the PPARC for support in the form of a
research studentship, postdoctoral, advanced and senior fellowships
respectively.
JPH acknowledges NASA grants NAG5-2523 and NAG5-4828.

\bibliographystyle{/opt/TeX/tex/bib/mn}

\bsp % ``This paper has been produced using the ...''
\label{lastpage}

\end{document}